\documentclass[graybox]{svmult}

\usepackage{mathptmx}       
\usepackage{helvet}         
\usepackage{courier}        
\usepackage{type1cm}        
\usepackage{amssymb}
\usepackage{amsmath}
\usepackage{amsfonts}
\usepackage{hyperref}
\usepackage{pictexwd,dcpic}
\usepackage{makeidx}         
\usepackage{graphicx}        
\usepackage{multicol}        
\usepackage[bottom]{footmisc}

\makeindex             
                       
\def\tit{\textit}

\def\tb{\bar{t}}
\def\ah{\hat{\alpha}}

\def\beq{\begin{equation}}
\def\eeq{\end{equation}}
\def\bea{\begin{eqnarray}}
\def\eea{\end{eqnarray}}
\def\ben{\begin{enumerate}}
\def\een{\end{enumerate}}
\def\la{\langle}
\def\ra{\rangle}
\def\a{\alpha}
\def\b{\beta}
\def\g{\gamma}\def\G{\Gamma}
\def\d{\delta}\def\D{\Delta}

\def\k{\kappa}
\def\l{\lambda}
\def\m{\mu}
\def\n{\nu}
\def\o{\omega}

\def\s{\sigma}
\def\t{\tau}

\def\del{\partial}

\def\half{{\textstyle{\frac{1}{2}}}}

\def\ut{\tilde{u}}

\def\vphi{\varphi}

\begin{document}

\title*{Black holes and Hawking radiation in spacetime and its analogues}
\author{Ted Jacobson}
\institute{Ted Jacobson \at Center for Fundamental Physics, Department of Physics, University of Maryland, College Park, MD 20742-4111, USA. \email{jacobson@umd.edu}}
%
%
\maketitle

\abstract{These notes introduce the fundamentals of black hole geometry, the thermality of the
vacuum, and the Hawking effect, in spacetime and its analogues. 
Stimulated emission of Hawking radiation, the trans-Planckian question,
short wavelength dispersion, and white hole radiation in the setting of analogue models
are also discussed. No prior knowledge of differential geometry, general relativity, or 
quantum field theory in curved spacetime is assumed. The discussion 
attempts to capture the essence of these topics without oversimplification.}

\newpage

\section{Spacetime geometry and black holes}
\label{sec:1}

In this section I explain how black holes are described in general relativity, starting with 
the example of a spherical black hole, and followed by the 1+1 dimensional generalization that 
figures in many analogue models. Next I discuss how symmetries and conservation
laws are formulated in this setting, and how negative energy states arise. 
Finally, I introduce the concepts of Killing horizon and surface
gravity, and illustrate them with the \tit{Rindler} or \tit{acceleration horizon},
which forms the template for all horizons.

\subsection{Spacetime geometry}
\begin{figure}[!t]
\begin{center}
\includegraphics[width=.25\textwidth]{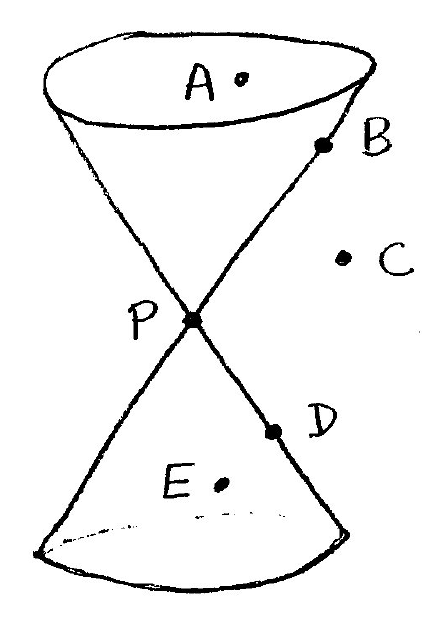}
\caption{\label{fig:lightcone} The light cone at an event $p$. 
The event $A$ is
future timelike related to $p$, while $B$, $C$, $D$, 
and $E$ respectively are future lightlike, spacelike,
past lightlike, and past timelike related to $p$.  
}
\end{center}
\end{figure}

The \textit{line element} or \tit{metric} 
$ds^2$ assigns a number to any infinitesimal displacement in spacetime. 
In a flat spacetime in a Minkowski coordinate system it takes the form
\beq\label{Mink}
ds^2 = c^2 dt^2 - (dx^2+dy^2+dz^2),
\eeq
where $t$ is the time coordinate, $x,y,z$ are the spatial Cartesian coordinates, and 
$c$ is the speed of light. Hereafter I will mostly employ units with $c=1$ except when
discussing analogue models (for which $c$ may depend on position and time when
using the Newtonian $t$ coordinate) .
When $ds^2=0$ the
displacement is called \textit{lightlike}, or \tit{null}. 
The set of such displacements at each event $p$ forms a double cone
with vertex at $p$ and spherical cross sections, called the \tit{light cone} or \tit{null cone}
(see Fig.~\ref{fig:lightcone}). 
Events outside the light cone are 
\tit{spacelike} related to $p$, while events inside the cone are either 
\tit{future timelike} or \tit{past timelike} related to $p$.
For timelike displacements, $ds^2$ determines the square of the 
corresponding \tit{proper time interval}.

The metric also defines the spacetime inner product 
$g(v,w)$ between two 4-vectors $v$ and $w$,
that is, 
\beq\label{MinkTensor}
g(v,w)=ds^2(v,w) = c^2 dt(v)dt(w) - [dx(v)dx(w)+dy(v)dy(w)+dz(v)dz(w)].
\eeq
Here 
$dt(v)=v^a\partial_a t = v^t$ is the rate of change of the $t$ coordinate along $v$, etc.

In a general curved spacetime the metric takes the form 
\beq\label{g}
ds^2 = g_{\a\b}(x) dx^\a dx^\b,
\eeq
where $\{x^\a\}$ are coordinates that label the points in a patch of a spacetime
(perhaps the whole spacetime), and there is an implicit summation
over the values of the indices $\a$ and $\b$.
The \tit{metric components}
$g_{\a\b}$ are functions of the coordinates, denoted $x$ in (\ref{g}).
In order to define a metric with Minkowski signature, the matrix
$g_{\a\b}$ must have one positive and three negative eigenvalues at each point.
Then \tit{local inertial coordinates} can be chosen in the neighborhood 
any point $p$ such that (i) the metric has the Minkowski form (\ref{Mink}) at $p$ 
and (ii) the first partial derivatives of the metric vanish at $p$. In 
two spacetime dimensions there are 9  independent second partials of the 
metric at a point. These can be modified by a change of coordinates
$x^\m\rightarrow x^{\m'}$, 
but the relevant freedom resides in the 
third order Taylor expansion 
coefficients $(\partial^3 x'^\m/\partial x^\a\partial x^\b\partial x^\g)_p$,
of which only 8 are independent because of the symmetry of mixed partials. 
The discrepancy $9-8=1$ measures 
the number of independent second partials of the metric that cannot 
be set to zero at $p$, which is the same as the number of independent
components of the \tit{Riemann curvature tensor} at $p$. So a single
curvature scalar characterizes the curvature in a two dimensional spacetime.
In four dimensions the count is $100-80=20$.

\subsection{Spherical black hole}

The Einstein equation has a unique (up to coordinate changes) 
spherical solution in vacuum for each mass,
called the \tit{Schwarzschild spacetime}.\index{Schwarzschild spacetime}

\subsubsection{Schwarzschild coordinates}\index{Schwarzschild coordinates}

The line element in so-called \tit{Schwarzschild coordinates} is given by 
\beq\label{S}
ds^2 = \left(1-\frac{r_s}{r}\right)d\tb^2 -  \left(1-\frac{r_s}{r}\right)^{-1}dr^2 - r^2(d\theta^2 +\sin^2\theta\, d\phi^2).
\eeq
Here
$r_s=2GM/c^2$ is the \tit{Schwarzschild radius}, with $M$ is the mass, and $c$ is set to 1. 
Far from the black hole, $M$ determines
the force of attraction in the Newtonian limit, and $Mc^2$ is the total energy of the spacetime.

The spherical symmetry is manifest in the form of the line element. The coordinates
$\theta$ and $\phi$ are standard spherical coordinates, while
$r$ measures $1/2\pi$ times the circumference of a great
circle, or the square root of $1/4\pi$ times the area of a sphere. 
The value $r=r_s$ corresponds to the \tit{event horizon}\index{event horizon}, as will be explained, 
and the value $r=0$ is the ``center", where the gravitational tidal force 
(curvature of the spacetime) is infinite. 
Note that  $r$  should \tit{not} to be thought of as the radial distance to $r=0$.
That distance isn't well defined until a spacetime path is chosen. (A path 
at constant $\tb$ does not reach any $r<r_s$.)

The coordinate $\tb$ is the \tit{Schwarzschild time}. It measures proper time at $r=\infty$, 
wheras at any other fixed $r$, $\theta$, $\phi$ the proper time interval is
$\D \t =\sqrt{1-r_s/r}\,  dt$. 
The coefficients in the line element are independent of $\tb$, hence the spacetime has
a symmetry under $\tb$ translation. This is ordinary time translation symmetry at
$r=\infty$, but it becomes a lightlike translation at $r=r_s$, 
and a space translation symmetry for $r<r_s$, since the coefficient
of $d\tb^2$ is negative there. The defining property of the Schwarzschild time coordinate,
other than that it measures proper time in the rest frame of the black hole at infinity, 
is that surfaces of constant $\tb$ are orthogonal, in the spacetime sense, to the 
direction of the time-translation symmetry, i.e.\ to the lines of constant 
$(r,\theta,\phi)$: there are no off-diagonal terms in the line element.
But this nice property is also why $\tb$ is ill behaved at the horizon.

\subparagraph{Redshift and horizon}

\begin{figure}[!t]
\begin{center}
\includegraphics[width=.375\textwidth]{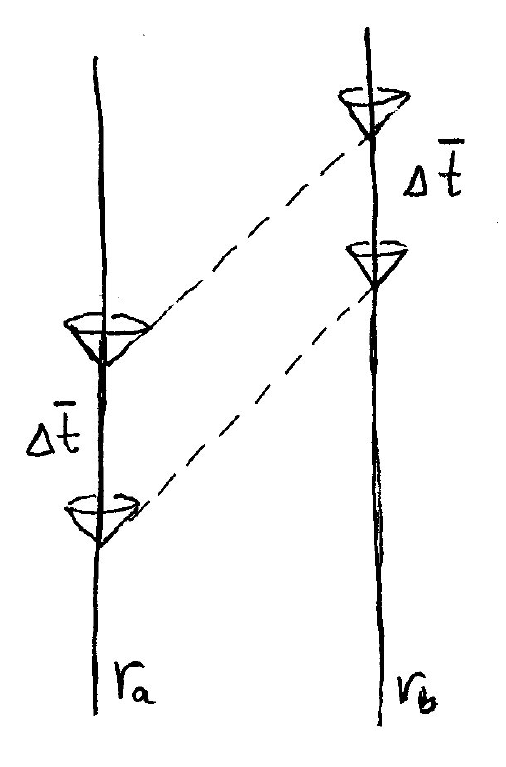}
\caption{\label{fig:redshift} Gravitational redshift.
Two lightrays propagating from $r_a$ to $r_b$, separated
by a coordinate time $\delta \bar{t}$. The corresponding
proper time at $r_a$ is less than that at $r_b$. 
}
\end{center}
\end{figure}

Suppose a light wave is generated with coordinate period $\D \tb$ at some radius $r_a$,
and propagates to another radius $r_b$ (see Fig.~\ref{fig:redshift}). 
Because of the time translation symmetry of the
spacetime, the coordinate period of the wave at $r_b$ will also be $\D\tb$. The ratio
of the proper time periods will thus be $\D\t_a/\D\t_b = \sqrt{1-r_s/r_a}/\sqrt{1-r_s/r_b}$,
and the ratio of the frequencies will the the reciprocal. This is the
\tit{gravitational redshift}. Note that as $r_a\rightarrow r_s$, the redshift is infinite.
The infinite redshift surface $r=r_s$ of the spherical black hole is the 
(stationary) event horizon. The same is true of the 1+1 dimensional black holes
we focus on later in these notes.

\begin{figure}[!t]
\begin{center}
\includegraphics[width=.375\textwidth]{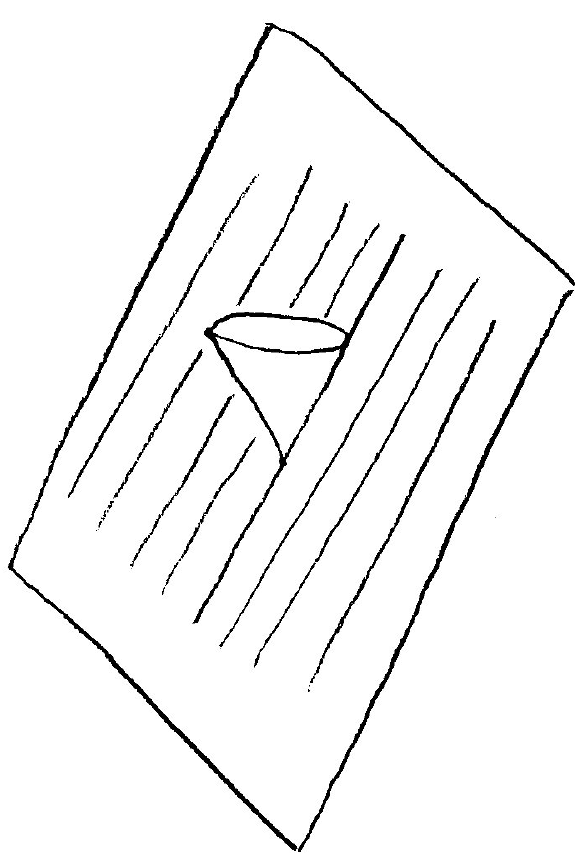}
\caption{\label{fig:nullsurface} A null surface is tangent 
to the local light cone. 
}
\end{center}
\end{figure}

It is worth emphasizing that for a \tit{non-spherical} 
stationary black hole, for instance a rotating black hole, 
the infinite redshift surface, where the time-translation 
symmetry becomes  lightlike, is generally \tit{not} the event horizon, because it is a
timelike surface. A timelike surface can be crossed in either direction. In order to 
be a horizon, a surface must be tangent to the local light cone at each point, so that it cannot
be crossed from inside to outside without going faster than light. At each point 
of such a \tit{null surface} there is one null tangent direction, and all other tangent 
directions are spacelike and orthogonal to the null direction (see Fig.~\ref{fig:nullsurface}). 
Therefore the 
null tangent direction is orthogonal to all directions in the surface, i.e. the null tangent is
also the normal. If the horizon
is a constant $r$ surface, then the gradient $\nabla_\a r$ is also orthogonal to 
all directions in the surface, so it must be parallel to the null normal. This means that
it is a null (co)vector, hence $g^{\a\b}\nabla_\a r \nabla_\b r=g^{rr}=0$ at the horizon.

\subsubsection{Painlev\'e-Gullstrand coordinates}\index{Painlev\'e-Gullstrand coordinates}

A new time coordinate $t$ that is well behaved at the horizon can be defined
by $t=\tb + h(r)$ for a suitable function $h(r)$ whose bad behavior at $r_s$ cancels that
of $\tb$. This property of course leaves a huge freedom in $h(r)$, but 
a particularly nice choice is defined by 
\beq\label{PG}
dt=d\tb + \frac{\sqrt{r}}{r-1} dr,\quad \mbox{i.e.}\quad t = \tb -2\sqrt{r}+\ln\left(\frac{\sqrt{r}+1}{\sqrt{r}-1}\right)
\eeq
where now I have adopted units with $r_s=1$. It is easy to see that the $t$-$r$ part
of the Schwarzschild line element takes the form
\bea
ds^2 &=& dt^2 -\left(dr +\sqrt{\frac1r}\, dt\right)^2- r^2(d\theta^2 +\sin^2\theta\, d\phi^2)\label{PGmetric}\\
&=& \left(1-\frac1r \right)dt^2 -\frac{2}{\sqrt{r}}dt\, dr - dr^2 - r^2(d\theta^2 +\sin^2\theta\, d\phi^2)\label{PGmetric2}
\eea
The new coordinate $t$ is called the \tit{Painlev\'e-Gullstrand} (PG) time.
At $r=1$ the metric coefficients are all regular, and indeed the coordinates are all well behaved there.
According to (\ref{PGmetric2}),  we have $ds^2=0$ along a line of constant $(r=1,\theta,\phi)$, so such 
a line is lightlike. Such lines generate the event horizon of the black hole. 
The PG time coordinate has some remarkable properties:
\begin{itemize}
\item the constant $t$ surfaces are flat, Euclidean spaces;

\item the radial worldlines 
orthogonal to the constant $t$ surfaces are timelike geodesics (free-fall trajectories)
along which $dt$ is the proper time.
\end{itemize}

For some practice in spacetime geometry, let me take you through verifying these properties.
Setting $dt=0$ in the line element we see immediately that $\{r,\theta,\phi\}$ are
standard spherical coordinates in Euclidean space. 
To find the direction orthogonal to a constant $t$ surface we could note that the gradient 
$\nabla_\a t$ has vanishing contraction with any vector tangent to this surface, which implies
that the contravariant vector $g^{\a\b}\nabla_\b t$, formed by contraction with the 
inverse metric $g^{\a\b}$, is orthogonal to the surface. Alternatively, we need not compute the
inverse metric, since the form of the line element (\ref{PGmetric}) allows us to read off the orthogonal
direction ``by inspection" as follows. 
Consider the inner product of two 4-vectors $v$ and $w$ in this metric, 
\bea
g(v,w) = &dt(v) dt(w)- \left(dr +\sqrt{\frac1r}\; dt\right)(v)\left(dr +\sqrt{\frac1r}\; dt\right)(w)\\ 
&- r^2d\theta(v) d\theta(w) -r^2 \sin^2\theta\, d\phi(v)d\phi(w),
\eea
using the notation of Eq. (\ref{MinkTensor}).
If the vector $v$ is tangent to the constant $t$ surface, then $dt(v)=0$, so the first term
vanishes. The remaining terms will vanish if $\left(dr +\sqrt{\frac1r}\; dt\right)(w)=d\theta(w)=d\phi(w)=0$.
Thus radial curves with $dr+\sqrt{1/r}\;dt=d\theta=d\phi=0$ are orthogonal to the surface, 
and along them $ds^2=dt^2$, i.e.\ $dt$ measures proper time along those curves.
Moreover, any other timelike 
curve connecting the same two spacetime points will have shorter proper time,
because the negative terms in $ds^2$ will contribute. The proper time is thus 
stationary with respect to first order variations of the curve, which is the defining property 
of a geodesic\index{geodesic}.\footnote{Even if the other terms in the line element  (\ref{PGmetric})  
had not been negative, they would not contribute to the first order variation 
in the proper time away from a path with $(dr +\sqrt{1/r}\; dt)=d\theta^2=d\phi^2=0$,
since the line element is quadratic in these terms. Thus the curve would still have been a geodesic
(although the metric signature would not be Lorentzian).}

\subsubsection{Spacetime diagram of the black hole}

\begin{figure}[!t]
\begin{center}
\includegraphics[width=.6\textwidth]{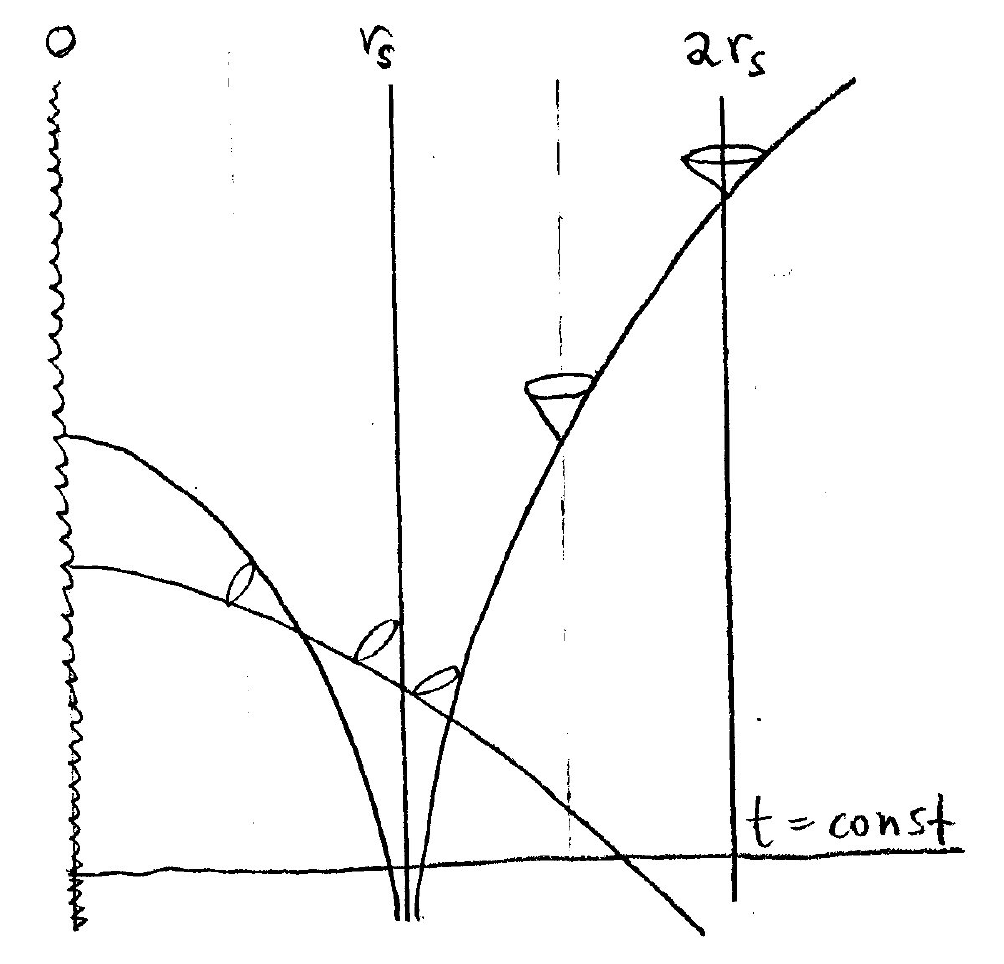}
\caption{\label{fig:PGdiagram} Painlev\'{e}-Gullstrand coordinate 
grid for Schwarzschild
black hole. Vertical lines have constant $r$, horizontal
lines have constant $t$. Shown are one ingoing radial 
light ray and three outgoing ones. The one outside the horizon
escapes to larger radii, the one on the horizon 
remains at $r_s$, and the one inside the horizon falls to 
smaller radii and into the singularity at $r=0$.  }
\end{center}
\end{figure}

The nature of the unusual geometry of the black hole spacetime can be grasped rather 
easily with the aid of a spacetime diagram (see Fig.~\ref{fig:PGdiagram}).
For the Schwarzschild black hole,
we may exploit the spherical symmetry and plot just a fixed value of the 
spherical angles $(\theta,\phi)$, and we may plot the lines of constant $r$ vertically
and the lines of constant PG time $t$ horizontally.  Then 
the time translation symmetry corresponds to
a vertical translation symmetry of the diagram.

The diagram comes alive when the light cones are plotted. At a given event, the
light cone is determined by $ds^2=0$, which for radial displacements corresponds
to the two slopes 
\beq\label{lightrays}
dt/dr = \frac{1}{\pm1-\sqrt{1/r}}\qquad \mbox{(radial lightrays)}
\eeq
Far from the horizon these are the outgoing and incoming lightrays $dt/dr\rightarrow\pm1$.
The ingoing slope is negative and gets smaller in absolute value as $r$ decreases, 
approaching $0$ as $r\rightarrow0$. 
The outgoing slope grows as $r$ decreases, until reaching infinity at the horizon at $r=1$. 
Inside the horizon it is negative, so an ``outgoing" lightray actually propagates to smaller values
of $r$. The outgoing slope also approaches $0$ as $r\rightarrow0$.

\subsubsection{Redshift of outgoing waves near the horizon}
\label{redshift}

An outgoing wave is stretched as it climbs away from the horizon. The lines of 
constant phase for an outgoing wave satisfying the relativistic wave equation
are just the outgoing lightrays (\ref{lightrays}). The rate of change of a wavelength
$\l$ is given by the difference of $dr/dt$ of the lightrays on the two ends of a wavelength, 
hence $d\l/dt = (d/dr)(dr/dt)\lambda$. The relative 
stretching rate is thus given by 
\beq\label{stretching}
\kappa\equiv\frac{d\l/dt}{\l}=\frac{d}{dr}\frac{dr}{dt}= \frac{c}{2r_s},
\eeq
where in the second step the expression is evaluated at the horizon, and 
the dimensionful constants are restored to better illustrate the meaning.
This rate is called the ``surface gravity"\index{surface gravity} 
$\kappa$ of the horizon. Later I will
explain different ways in which the surface gravity can defined and calculated.

We can go further and use the lightray equation (\ref{lightrays}) to obtain 
an approximate expression for the wave phase near the horizon.  
Consider an outgoing wave of the form $e^{i\phi}$, with 
$\phi = -\o t + \int^r k(r') dr'$. (This simple harmonic $t$ dependence is 
exact because the metric is independent of $t$.) Along 
an outgoing lightray the phase is constant: $0 = d \phi = -\o dt + k(r) dr$, 
so 
\beq
k(r) = \frac{\o}{1- r^{-1/2}} \sim \frac{2\o}{r-1} = \frac{\o/\k}{r-r_s}, 
\eeq
where in the second step
a near horizon approximation is used, and in the last step
the dimensionful constants are again restored.  The  
wave thus has the near-horizon form 
\beq\label{log}
e^{-i\o t}e^{i(\o/\k)\ln(r-r_s)}.
\eeq
Note that the surface gravity appears in a ratio with the 
wave frequency, and there is a logarithmic 
divergence in the outgoing wave phase at the horizon.

\subsection{Effective black hole and white hole spacetimes}

Many black hole analogues can be described with one spatial dimension, and I will
focus on those here. They are simple generalizations of the radial 
direction for a spherical black hole. 

Waves or quasiparticles in a stationary 1+1 dimensional setting
can often be described by a relativistic field in an effective spacetime 
defined by a metric of 
the form
\beq\label{1+1}
ds^2 = c(x)^2 dt^2 -[dx - v(x) dt]^2=  [c(x)^2- v(x)^2]dt^2 +2v(x) dt\, dx -dx^2.
\eeq
In fact, any stationary two dimensional metric can be put in this form,
with $c(x)=1$, by a 
suitable choice of coordinates (see e.g.\ Appendix A in Ref.~\cite{Corley:1997ef} for 
a proof of this statement). If $c(x)=1$ 
this corresponds to the PG metric, with 
$x\leftrightarrow r$ and $v(x)\leftrightarrow -1/\sqrt{r}$.
A horizon exists in the spacetime (\ref{1+1})
if $|v(x)| >|c(x)|$ somewhere. 


The metric (\ref{1+1}) would arise for example
in a Newtonian setting of a fluid, with velocity $v(x)$ in a ``laboratory frame",
with $c(x)=c$ a constant speed of sound.
In that example, the coordinate $x$ would measure distance in the lab at fixed Newtonian
time $t$, and the metric would describe the effective spacetime for 
waves in the fluid that propagate at speed 
$c$ relative to the local rest frame
of the fluid. If the wave speed 
in the frame of the medium depends on some ambient local conditions 
then $c(x)$ will depend on position. 

\subparagraph{Moving texture}\index{texture}
In some models
the medium may be at rest in the lab, but the local conditions
that determine the wave speed may depend on both time and 
space in a ``texture" that moves. (If the motion is uniform then
in the frame of the texture this is equivalent to the previous case.)
An example of a 
line element of this sort is $[c(y-wt)]^2\, dt^2 - dy^2$. Here 
again $y$ measures proper distance in the lab at Newtonian time $t$,
and the texture moves in the $y$ direction with constant speed $w$.
The line element may not look stationary, but it has a symmetry 
under $t\rightarrow t+\D t$ combined with $y\rightarrow y+w\D t$.

\subparagraph{Black hole -- white hole pair}\index{white hole}

\begin{figure}[!t]
\begin{center}
\includegraphics[width=.6\textwidth]{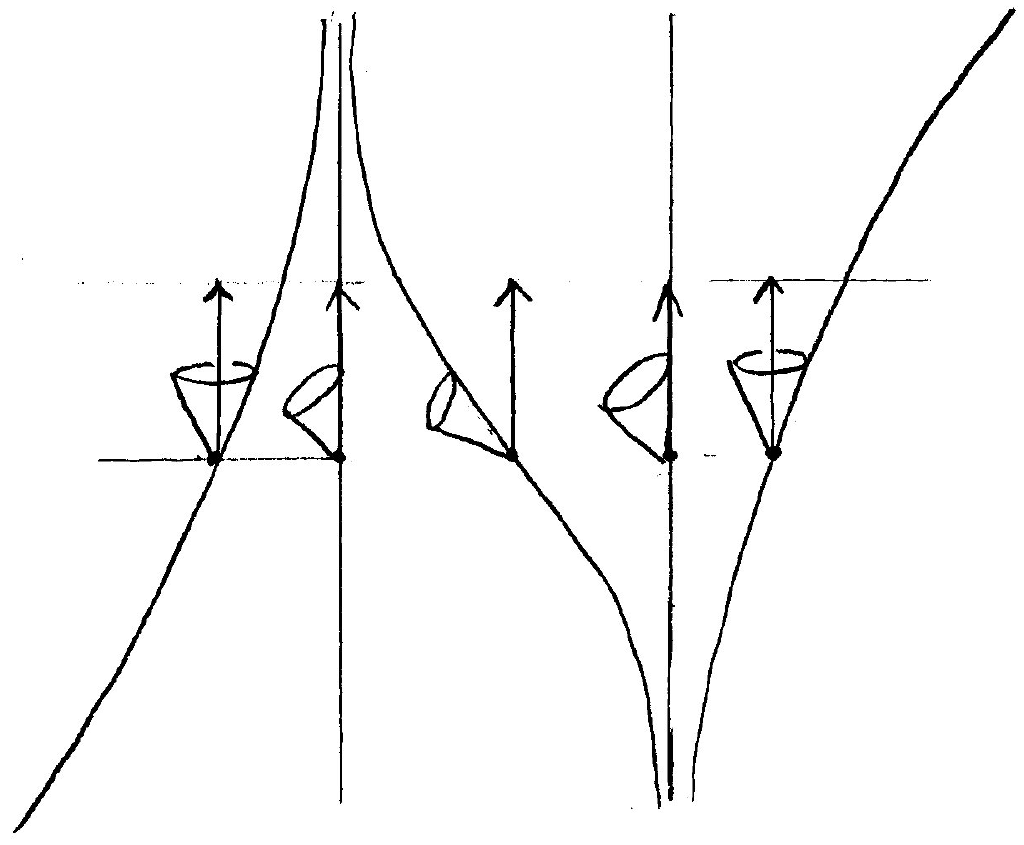}
\caption{\label{fig:bhwh} 
Black hole horizon on the right and white hole horizon 
on the left. The vertical arrows depict the Killing vector,
which is spacelike in the ergoregion between the horizons 
and timelike outside.}
\end{center}
\end{figure}

An example that often arises has
$v(x)<-c(x)<0$ in a finite interval $(x_-,x_+)$. Then 
$x_+$ is a black hole horizon, 
analogous to the one previously discussed for 
the PG spacetime, and $x_-$ is a \tit{white hole} horizon:
no waves can escape from the region $x<x_+$
into the region $x>x_+$, and no waves can enter the
region $x>x-$ from the region $x<x_-$. 
The region between the horizons
is of finite size and nonsingular. Fig.~\ref{fig:bhwh} 
is a spacetime diagram
of this scenario.
Black hole horizon on the right and white hole horizon 
on the left. The vertical arrows depict the Killing vector,
which is spacelike in the ergoregion between the horizons 
and timelike outside. 

\subsection{Symmetries, Killing vectors, and conserved quantities}\index{Killing vector}

Each symmetry of the background spacetime and fields
leads to a corresponding conservation law.
The most transparent situation is when the metric and 
any other background fields are simply independent
of some coordinate. This holds for example with the
Schwarzschild metric (\ref{S}), which is independent of both $t$ and 
$\phi$. Of course the spherical symmetry goes beyond
just $\phi$ translations, but the other rotational symmetries 
are not manifest in this particular form of the line element. 
They could be made manifest by a change of coordinates however,
but not all at once.
To be able to talk about 
symmetries in a way that is independent of whether or not
they are manifest it is useful to introduce the notion of a
\tit{Killing vector field}. 
The flow of the spacetime along the 
integral curves of a Killing vector is a symmetry of the spacetime.

Suppose translation by 
some particular coordinate $x^{\ah}$ ($\ah$ indicates 
one particular value of the index $\a$) is a manifest
symmetry. 
The metric components satisfy 
$g_{\m\n,\ah}=0$, where the comma notation
denotes partial derivative with respect to $x^{\ah}$.
The  corresponding Killing vector, written
in these coordinates, is $\chi^\m=\d^\m_{\ah}$, i.e.\
the vector with all components 
zero except the $\ah$ component which is 1.
Then the symmetry is expressed by the 
equation $g_{\m\n,\a}\chi^\a=0$. 
This holds only in special coordinate systems
adapted to the Killing vector. It
is not a tensor equation, since the 
partial derivative of the metric is not a tensor.

It may be helpful to understand 
that this condition is equivalent to 
the covariant, tensor equation for a 
Killing vector, 
\beq\label{Keq}
\chi_{\a;\b}+\chi_{\b;\a}=0,
\eeq
where the semicolon denotes the covariant
derivative. This is called \tit{Killing's equation}.
One way to see the equivalence is to use the fact
that in a local inertial coordinate system at a
point $p$, the covariant derivative reduces to the
partial derivative, and the partials of the metric are zero.
Thus Killing's equation at the point $p$ becomes 
$\chi^\s{}_{,\b}\eta_{\s\a}+\chi^\s{}_{,\a}\eta_{\s\b}=0$,
where $\eta_{\s\t}$ is the Minkowski metric. 
This implies that the infinitesimal flow
$x^\s\rightarrow x^\s + \epsilon \chi^\s(x)$ 
generated by $\chi^\a$ is, to lowest order, 
a translation plus a Lorentz 
transformation, i.e.\ a symmetry of the metric.\footnote{For a more
computational proof, note that since Killing's equation is a tensor
equation it holds in all coordinate systems 
if it holds in one. In a coordinate system for which 
$\chi^\m=\d^\m_{\ah}$
we have 
$\chi_{\a;\b}= g_{\a\m}\chi^\m{}_{;\b} 
= g_{\a\m}\G^\m{}_{\b\s}\chi^\s
= \half(g_{\a\b,{\s}}+g_{\a{\s},\b}-g_{\b{\s},\a})\chi^\s$.
If $\chi^{{\s}}$ is a Killing vector the first term vanishes
in this adapted coordinate system, and the
remaining expression is antisymmetric in $\a$ and $\b$,
so adding $\chi_{\b;\a}$ yields zero. 
Conversely, if Killing's equation holds, 
the entire expression is antisymmetric
in $\a$ and $\b$, so the first term must vanish.}

A simple example is the Euclidean plane with line element
$ds^2 = dx^2 + dy^2 = dr^2 + r^2 d\phi^2$ in Cartesian
and polar coordinates respectively. The rotation Killing vector 
about the origin in polar coordinates is just $\partial_\phi$, 
with components $\d^\a_\phi$, as the metric components are
independent of $\phi$. The same Killing vector in 
Cartesian coordinates is $x\partial_y - y\partial_x$. This
satisfies Killing's equation since 
$\chi_{x,x}=0=\chi_{y,y}$, and 
$\chi_{x,y}+\chi_{y,x}=-1+1=0$.

\subsubsection{Ergoregions}\index{ergoregion}

It is of paramount importance in
black hole physics that a Killing field may be timelike in 
some regions and spacelike in other regions of a spacetime.
For example in the Schwarzschild spacetime, say in PG coordinates
(\ref{PGmetric}), or the 1+1 dimensional generalization
(\ref{1+1}) the Killing vector $\partial_t$ is timelike outside the horizon, 
but it is lightlike on the horizon and 
spacelike inside. For the black hole-white hole pair discussed above, 
it is the region between the black and white hole horizons (see Fig.~\ref{fig:bhwh}).
This is evident because the coefficient of $dt^2$ in the line
element becomes negative.

A region where an otherwise timelike
Killing vector becomes spacelike is called an \tit{ergoregion}. 
(The reason for the name will become clear below.)
The boundary of this region is called the \tit{ergosurface},\index{ergosurface}
and it is a surface of infinite redshift, since the norm of the
time translation Killing vector vanishes there.
An ergoregion need not lie behind a horizon. For instance it 
occurs outside the horizon (as well as inside) of
a spinning black hole. In analogue models, ergoregions can arise for example around 
a vortex~\cite{Barcelo:2005fc} 
or in a moving soliton in superfluid ${}^3$He-A~\cite{Jacobson:1998ms}. 
For the Schwarzschild 
black hole, and the 1+1 dimensional generalization (\ref{1+1}), however,
the ergoregion always corresponds to the region inside the horizon.

\subsubsection{Conserved quantities}

Particle trajectories (both timelike and lightlike) can be 
determined by the variational principle
$\d\int L\, d\l=0$ with Lagrangian 
$L= \half g_{\m\n}(x)\dot{x}^\m\dot{x}^\n$.
Here $\l$ is a path parameter and the dot denotes
$d/d\l$. The Euler-Lagrange equation is the geodesic
equation for motion in the metric $g_{\m\n}$ with affine parameter
$\l$. If the metric is independent of $x^{\ah}$ then the 
corresponding conjugate momentum 
$p_{\ah}=\partial L/\partial\dot{x}^{\ah}= g_{\m\ah}\dot{x}^\m$
is a constant of motion. Note that this
momentum can also be expressed as
the inner product of the 4-velocity $u^\n=\dot{x}^\n$ 
with the Killing field, $u\cdot\chi= g_{\m\n}\dot{x}^\m\chi^\n=g_{\m\ah}\dot{x}^\m$.

\subparagraph{Killing energy and ergoregions}

The conserved momentum conjugate to a
particular timelike Killing field is called \tit{Killing energy}.\index{Killing energy} 
For a particle with rest mass $m$, the physical 
4-momentum would be $p=mu$, so the Killing energy as
defined above is actually the Killing energy per unit rest mass. 
For a massless particle, the physical 4-momentum is proportional
to the lightlike 4-velocity, scaled so that the time component 
in a given frame is the energy in that frame. 
In both cases, the true Killing energy is the 
inner product of the 4-momentum and the Killing vector,
\beq
E_{\rm {\scriptstyle{Killing}}}=p\cdot\chi.
\eeq
The 4-momentum of a massive particle is timelike, while
that of a massless particle is lightlike. In both cases,
for a physical state (i.e. an allowable excitation of the vacuum), 
stability of the local vacuum implies that 
the energy of the particle is positive 
as measured locally in any rest frame.
This is equivalent to the statement that 
$p$ is a \tit{future pointing} 4-vector. 

\begin{figure}[!t]
\begin{center}
\includegraphics[width=.45\textwidth]{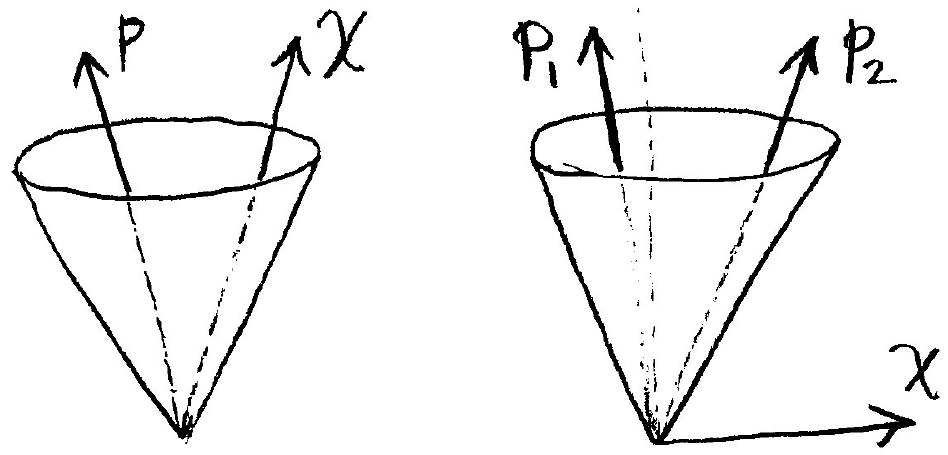}
\caption{\label{fig:KillingEnergy} Killing energy. 
On the left the Killing vector $\chi$ is timelike, hence all 
future causal (timelike or lightlike) 4-momenta have positive
$\chi$-energy. On the right $\chi$ is spacelike, hence future 
causal 4-momenta like $p_2$ can have negative $\chi$-energy,
while others like $p_1$ have positive $\chi$-energy.}
\end{center}
\end{figure}

The importance of ergoregions stems from the fact that 
negative Killing energy physical states exist there.
This happens because a future pointing 4-momentum can 
of course have a negative inner product with a spacelike 
vector (see Fig.~\ref{fig:KillingEnergy}). 
In an ergoregion, the Killing energy is what
would normally be called a linear momentum component,
and there is of course no lower limit on the linear momentum
of a physical state. 

Penrose~\cite{Penrose:1969pc,Penrose:1971uk} 
realized that the existence of an ergoregion outside 
a spinning black hole implies that energy can be extracted
from the black hole by a classical process, 
at the cost of lowering the angular momentum.
This is the \tit{Penrose process},\index{Penrose process} 
whose existence led to the
discovery of black hole thermodynamics. 
For a non-spinning black hole the ergoregion lies 
inside the horizon, so no classical process can 
exploit it to extract energy, but the \tit{Hawking
effect} is a quantum process by which energy is 
extracted.

\begin{figure}[!t]
\begin{center}
\includegraphics[width=.375\textwidth]{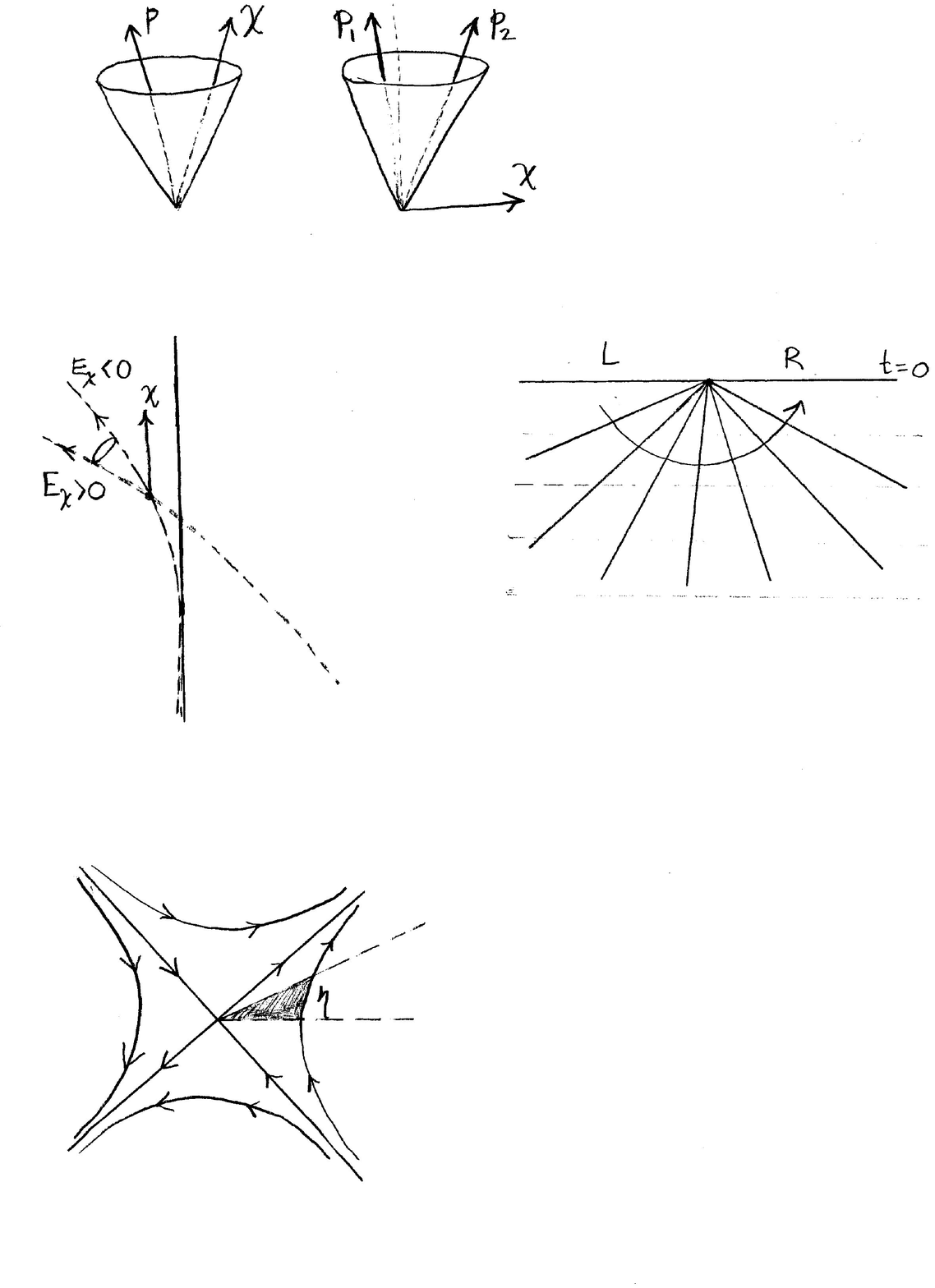}
\caption{\label{fig:outgoing} Inside the horizon, 
the Killing vector $\chi$ is spacelike, outgoing
radial particles have negative $\chi$-energy, and infalling ones
have positive $\chi$-energy. (Since the latter come from 
outside the ergorgion, and Killing energy is conserved, they
must have positive Killing energy.) 
}
\end{center}
\end{figure}

What do the negative Killing energy states ``look like"?
A particle with negative Killing energy cannot
escape from the ergoregion, nor can it have fallen freely 
into the ergoregion, because Killing energy is conserved
along a geodesic and 
it must have positive Killing energy if outside the ergoregion.
For example, in the 1+1 black hole, or in the radial direction
of the Schwarzschild solution, a massless particle
with negative Killing energy
inside the horizon 
must be ``outgoing" as seen by a local 
observer (see Fig.~\ref{fig:outgoing}).

\subsection{Killing horizons and surface gravity}\label{sec:KillingH}

An event horizon can be defined purely in terms of the causal structure
of a spacetime, and is meaningful even when the spacetime is
not stationary, i.e.\ has no time translation symmetry. 
A \tit{Killing horizon} on the other hand 
is a lightlike hypersurface (surface of one less dimension than the 
whole spacetime) generated by the 
flow of a Killing vector. This is sometimes called the
\tit{horizon generating Killing vector}.\index{horizon generating Killing vector}

The Schwarzschild event horizon is 
a Killing horizon with respect to the Killing vector $\partial_t$,
as is the horizon of the 1+1 black hole. A distinction arises
in the case of a stationary black hole with spin. Then the 
Killing vector $\partial_t$ that is a time translation at spatial infinity
becomes lightlike at the boundary of the ergoregion, which 
lies outside the event horizon. However 
that boundary is timelike, so the ergosurface is not a Killing horizon.
The event horizon of a spinning black hole is nevertheless
a Killing horizon, but for a Killing vector $\partial_t +\Omega_H \partial_\phi$ 
that is a linear combination of the time translation
and rotation Killing vectors, $\Omega_H$ being the angular velocity 
of the horizon.  In the effective spacetime of a moving texture
in superfluid ${}^3$He-A, the horizon generating Killing vector
has the similar form $\partial_t +w \partial_x$, where $\partial_t$
and $\partial_x$ are time and space translation Killing vectors,
and the constant $w$ can be thought of as the transverse velocity
of the horizon~\cite{Jacobson:1998ms}.

\subparagraph{Rindler (acceleration) horizon}

\begin{figure}[!t]
\begin{center}
\includegraphics[width=.45\textwidth]{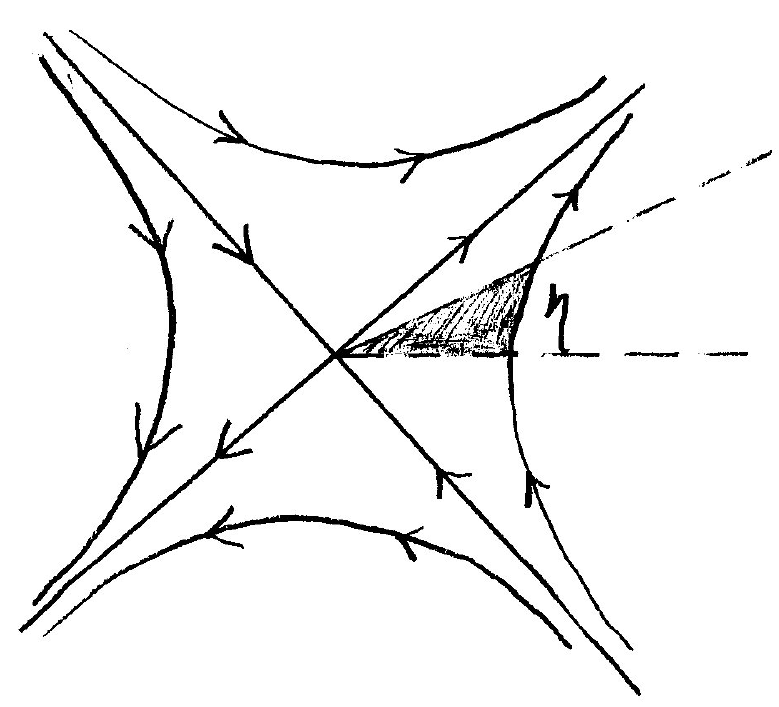}
\caption{\label{fig:boost} Boost killing flow in Minkowski space (\ref{MinkPolar}). 
Curves of constant
$\ell$ are hyperbolic flow lines. Lines of constant $\eta$ are radial
from the origin, and $\eta$ measures the hyperbolic opening angle of the 
shaded wedge.
}
\end{center}
\end{figure}

A simple yet canonical example of a Killing horizon is the Rindler horizon\index{Rindler horizon} 
in Minkowski spacetime. The relevant Killing symmetry here is Lorentz boosts\index{Lorentz boost} is a certain direction.
Geometrically, these are just hyperbolic rotations.\index{hyperbolic rotation} 
For example, using the 
Minkowski coordinates of (\ref{Mink}) a \tit{boost Killing vector}\index{boost Killing vector} is 
\beq
\chi_B=x\partial_t+t\partial_x.
\eeq
This has covariant components
$(\chi_B)_\a = \eta_{\a\b}\chi_B^\b=(x, -t)$ and so obviously satisfies Killing's
equation (\ref{Keq}). It can also be made manifest by changing from Minkowski
to polar coordinates: 
\beq\label{MinkPolar}
dt^2 - dx^2 = \ell^2 d\eta^2 -d\ell^2 .
\eeq
Then the 
boost symmetry is just rotation of the hyperbolic angle\index{hyperbolic angle}
$\eta$, i.e.\
\beq
\chi_B = \partial_\eta.
\eeq
The flow lines of the Killing field are hyperbolas (see Fig.~\ref{fig:boost}).
Note that the polar coordinate system covers only
one ``Rindler wedge",\index{Rindler wedge}
e.g. $x>|t|$ of the Minkowski spacetime. The full Killing horizon
is the set $|x|=|t|$.

\subsubsection{Surface gravity}\label{sec:kappa}\index{surface gravity}
Associated to a Killing horizon is a quantity $\kappa$ called the \tit{surface gravity}.
There are many ways to define, calculate, and think of the surface gravity. 
It was already introduced in Sec. \ref{redshift}, as the relative rate of stretching of
outgoing wavelengths near the horizon. 
I will mention here several other definitions, which are given directly in terms of the
geometry of the horizon.

Geometrically, the simplest definition of surface gravity may be via 
\beq
[\chi_{[\a,\b]}\chi^{[\a,\b]}]_H = -2\k^2,
\eeq
horizon the square bracket on indices denotes antisymmetrization,
and the subscript $H$ indicates that the quantity is evaluated on the horizon.
That is, $\k$ is the magnitude of the infinitesimal Lorentz transformation
generator. However the meaning of this is probably not very intuitive.

The conceptually simplest definition might be 
the rate at which the norm of the Killing vector vanishes
as the horizon is approached from outside. That is, 
\beq\label{kappa1}
\k=||\chi|_{,\a}|_H,
\eeq
the 
horizon limit of the norm of the gradient of the norm of $\chi$.
Notice that if the Killing vector is rescaled by a constant multiple
$\chi\rightarrow \a\chi$, then it remains a Killing vector, and the 
surface gravity for this new Killing vector is $\a\kappa$. 
This illustrates the important point that the intrinsic structure of 
a Killing horizon alone does not suffice to define the surface
gravity. Rather, a particular normalization of the Killing vector is required.
The symmetry implies that $\k$ is constant along a particular
null generator of the horizon, but in general it need not be the same on all
generators. For a discussion of conditions under which the 
surface gravity can be proved to be constant see \cite{Wald:1999vt}.

The surface gravity (\ref{kappa1}) has the interesting property that it is
{\it conformally invariant}. That is, it is
unchanged by a conformal rescaling of the metric
$g_{ab}\rightarrow \Omega^2g_{ab}$, provided the conformal
factor $\Omega$ is regular at the horizon~\cite{Jacobson:1993pf}. 
This follows simply because 
$|\chi|$ is rescaled by $\Omega$, while the norm of its gradient is
rescaled by $\Omega^{-1}$, and the contribution from $d\Omega$ vanishes
since it is multipled by $|\chi|_H$ which vanishes.

For the metric (\ref{1+1}) and the Killing vector
$\chi=\partial_t$ we have $|\chi|=\sqrt{c^2-v^2}$, 
which depends on $x$ and not $t$.  Thus
$\k=(-g^{xx}\partial_x|\chi|\, \partial_x|\chi|)^{1/2}_H$,
and the minus sign arises because the gradient is spacelike outside the
horizon. At a horizon where $v=c$ this evaluates to 
$|\partial_x (v-c)|_H$, while at a horizon where $v=-c$ it would instead be
$|\partial_x (v+c)|_H$.

In case $c=constant$, the surface gravity is thus just the gradient of the
flow speed at the horizon. 
A covariant and more general
version of this can be formulated. 
Any observer falling freely across a horizon can define the velocity 
of the static frame relative to himself, and can evaluate the
spatial gradient of this velocity in his frame. If he has unit
Killing energy ($u\cdot\chi=1$) then it can be shown that 
this gradient, evaluated at the horizon, 
agrees with the surface gravity~\cite{Jacobson:2008cx}.
Another interesting observation is that this velocity gradient
has a sort of ``cosmological"
interpretation as the local fractional rate of expansion (``Hubble constant")
of the distances separating a family of freely falling observers 
stretched along the direction of the Killing frame velocity~\cite{Jacobson:2008cx}.
At the horizon, for unit energy observers, this expansion rate is 
the same as the surface gravity.

Computationally, a somewhat simpler definition of surface gravity is
via 
\beq\label{kappa2}
[\partial_\a(\chi^2)=-2\k\chi_\a]_H.
\eeq
This is at least well-defined: since $\chi^2$ vanishes
everywhere on the Killing horizon, its gradient has zero contraction
with all vectors tangent to the horizon. The same is true for 
$\chi_a=g_{\a\b}\chi^\b$, so these two co-vectors must be parallel.
If using a coordinate component of  
this equation to evaluate $\kappa$, it is important that
the coordinate system be regular at the horizon. 
For the metric (\ref{1+1}), we may just evaluate the 
$x$ component of this equation:
$\partial_x(c^2-v^2)=-2\k\chi_x=-2\k g_{xt}=-2\k v$,
which on the horizon $v=c$ 
yields $\k = [\partial_x (v-c)]_H$ as before. (Note that this
definition does not come with an absolute value. At a horizon
$v=-c$ it yields $\k = [\partial_x (v+c)]_H$.)


\subparagraph{Surface gravity of the Rindler horizon}
 
The surface gravity of the Rindler horizon 
can be computed for example using the
polar coordinates to evaluate (\ref{kappa1}). Then the norm
of the Killing vector is just $\ell$, so 
$\partial_\a|\chi_B|=\d_\a^\ell$, which has norm 1.
Thus the boost Killing vector has unit surface gravity. 
Alternatively, we may use the $x$ component of (\ref{kappa2}):
$\partial_x\chi_B^2=x^2-t^2= 2x$, and $-2\k(\chi_B)_x = 2\k t$, so 
$\k= (x/t)_H = \pm 1$ On the future horizon $x=t$ and this is 
positive, while on the past horizon it is negative. Usually
one is only interested in the absolute value. 

Finally, it is sometimes of interest to use the proper time
along a particular hyperbola rather than the hyperbolic angle
as the coordinate. On the hyperbola located at $\ell=\ell_0$
the proper time is $d\t=\ell_0 d\eta$. The Minkowski
line element can be written in terms of the time coordinate $\t=\ell_0\eta$
as $ds^2 = (\ell/\ell_0)^2 d\t^2 -d\ell^2$. The scaling of the Killing field
$\partial_\t=(1/\ell_0)\partial_\eta$ that generates proper time
flow on this particular hyperbola has surface gravity $\k=1/\ell_0$.
This is also equal to the \tit{acceleration} of the 
hyperbolic worldline. The relation between the surface gravity and
acceleration can be shown quite generally using coordinate free
methods, but here let's just show it by direct computation
using Cartesian coordinates. The 4-velocity of the hyperbola
is the unit vector $u=\ell_0^{-1}(x,t,0,0)$, and the acceleration
of this worldline is $(u\cdot\nabla) u= \ell^{-2}_0(x\partial_t + t\partial_x)(x,t,0,0)
 =\ell^{-2}_0(t,x,0,0)$. The norm of the spacelike vector 
 $(t,x,0,0)$ is $\ell_0$, so the norm
 of the acceleration is $1/\ell_0$.

\section{Thermality of the vacuum}
\label{sec:2}

The subject of the rest of these notes is the Hawking effect,
i.e.\  the emission of thermal radiation from a black hole.
The root of the Hawking effect is the thermality of 
the vacuum in flat spacetime. 
This thermality is known as the Unruh,\index{Unruh effect}  
or Fulling-Davies-Unruh, effect~\cite{Crispino:2007eb}. 
In its narrowest form,
this is the fact that 
a probe with uniform proper acceleration $a$, moving 
through the vacuum of a quantum field in flat spacetime, 
is thermally excited at the Unruh temperature 
\beq\label{T_U}
T_U=\hbar a/2\pi c.
\eeq
(I've restored $c$ here to show where it enters, but will
immediately revert to units with $c=1$.)
When described this way, however, too much attention is focused on the 
probe and its acceleration. 

Underlying the response of the probe is a 
rather amazing general fact:\index{amazing fact} 
when restricted to a Rindler wedge, 
the vacuum of a relativistic quantum field is
a canonical thermal state with density matrix 
\beq\label{rhoR}
\rho_R\propto\exp(-2\pi H_\eta/\hbar),
\eeq
where $H_\eta$ is the ``boost Hamiltonian"\index{boost Hamiltonian} or 
``Rindler hamiltonian"\index{Rindler Hamiltonian} generating shifts of the 
hyperbolic angle coordinate $\eta$ defined in (\ref{MinkPolar}). 
In terms of Minkowski coordinates
$(t,x,y,z)$, $H_\eta$ is given on a $t=0$ surface of the Rindler wedge 
by
\beq
H_\eta=\int_{\Sigma_R} T_{ab}\chi_B^a d\Sigma^b = \int x\, T_{tt}\, dxdydz,
\eeq
where $T_{ab}$ is the energy-momentum tensor.
The ``temperature" of the thermal state (\ref{rhoR}) is 
\beq\label{TR}
T_R=\hbar/2\pi.
\eeq
Like a rotation angle, the 
hyperbolic angle is dimensionless, so the 
boost generator and temperature have dimensions of angular momentum.

Note that the thermal nature of the vacuum in the 
wedge does not refer to any particular
acceleration, and it characterizes the state even on a single 
time slice. Nevertheless it does directly predict the Unruh effect.
A localized probe that moves along a particular hyperbolic trajectory
at proper distance $\ell_0$ from the vertex of the wedge
has proper time interval $d\t = \ell_0  d\eta$ (cf.~\ref{MinkPolar}). When scaled
to generate translations of this proper time 
the field Hamiltonian is thus $H_\t=\ell_0^{-1}H_\eta$,
and the corresponding temperature is 
$T_0=\ell_0^{-1}\hbar/2\pi$. The proper acceleration 
of that hyperbola is $\ell_0^{-1}$, so the probe will be excited at
the Unruh temperature (\ref{T_U}).\index{Unruh temperature} 

The thermality of the vacuum in one wedge
is related to entanglement between the quantum states
in the right and left wedges. 
It can be understood using a simple, but abstract
and formal, argument that employs
the path integral expression for the ground state.
Because the result is so central to the subject, 
I think this argument deserves to be explained. 

The vacuum $|0\ra$  
is the ground state of the field Hamiltonian $H$, 
and can therefore be projected out of any state $|\chi\ra$
as $|0\ra\propto\lim_{t\rightarrow\infty}e^{-tH}|\chi\ra$, as long as 
$\la0|\chi\ra\ne0$. The
operator $e^{-tH}$ can be 
thought of as the 
time evolution operator for an imaginary time $-it$,
and its matrix elements can be represented by a
path integral over fields $\phi$ on Euclidean space. This yields
a path integral representation for the vacuum wave functional,\index{vacuum wave functional}
\beq\label{pathintegral}
\Psi_0[\phi]\propto 
\lim_{t\rightarrow\infty}\la\phi|e^{-tH}|\chi\ra\propto \int_{\phi(t=-\infty)=\chi}^{\phi(t=0)=\phi} {\cal D}\phi\, e^{-S/\hbar},
\eeq
where $S$ is the Euclidean action corresponding to the Hamiltonian $H$.
The standard demonstration of this path integral expression for matrix elements
of $e^{-tH}$ proceeds by slicing
the Euclidean space into steps of constant Euclidean time, and exploits
the time translation invariance of the Hamiltonian. If the original Hamiltonian is
also Lorentz boost invariant, then the Euclidean action is also rotationally invariant.
This extra symmetry leads to an alternate interpretation of the path integral.

\begin{figure}[!t]
\begin{center}
\includegraphics[width=.6\textwidth]{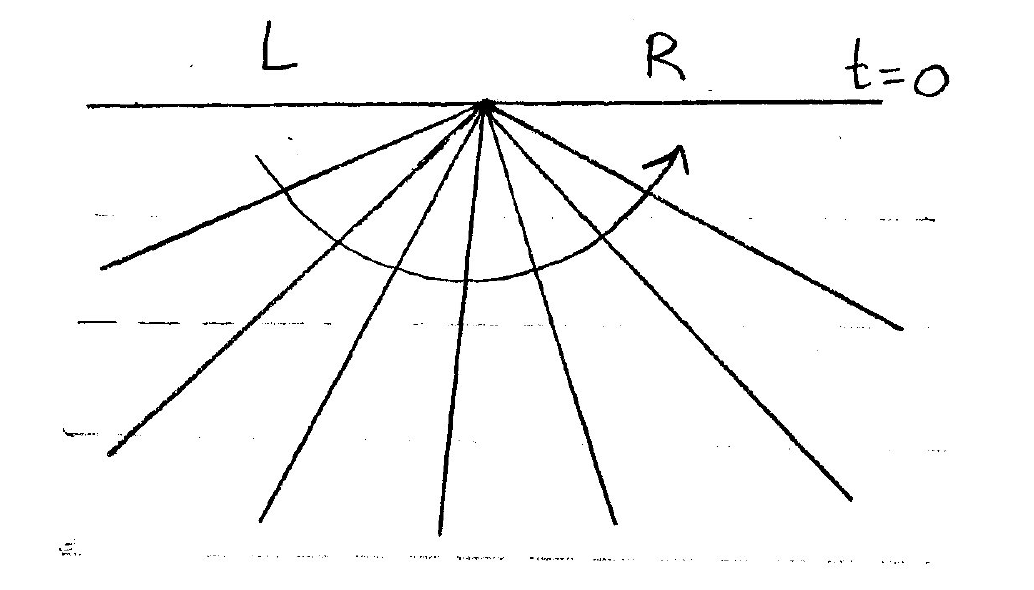}
\caption{\label{fig:slicings} Euclidean Minkowski space with boundary at $t=0$. 
When the path integral (\ref{pathintegral}) is sliced by constant 
$t$ surfaces it presents the vacuum wave-functional.
When sliced by constant angle surfaces, it presents
matrix elements of the operator $\exp(-\pi H_\eta)$,
where $H_\eta$ is the Lorentz boost generator.}
\end{center}
\end{figure}

Fixing a particular 
rotational symmetry, e.g. around the origin in the 
Euclidean $tx$ plane, 
we may choose to slice the Euclidean space into steps of constant angle
around the corresponding vertex (see Fig.~\ref{fig:slicings}). 
This vertex divides the time slice $t=0$ into two halves, 
and the final field configuration $\phi$ restricts to some
$\phi_L$ and $\phi_R$ on the left and right sides respectively.
These configurations
define Dirac ``bras" $\la\phi_L|$ and $\la \phi_R|$ in the 
duals of the left and right side Hilbert spaces ${\cal H}_L$ and ${\cal H}_R$.
The full Hilbert space is the tensor product ${\cal H}_L\otimes {\cal H}_R$.

With this angular slicing, (and not worrying about boundary conditions at the vertex), 
we can think of the path integral as producing the matrix element of the
operator $\exp(-\pi H_\eta)$ between $\phi_L$, regarded now as an \tit{initial} state,
and the final state $\phi_R$,
\beq\label{Psi0}
\Psi_0[\phi_L,\phi_R]\propto \la\phi_R|e^{-\pi H_{\eta}}J|\phi_L\ra.
\eeq
Here $H_{\eta}$ is the boost Hamiltonian, which is the generator
of angle shifts, 
and $\pi$ is the rotation angle in the Euclidean plane.
(The rotation angle is to the boost angle as the 
Euclidean time is to the Minkowski time.) 
The final state bra $\la \phi_L|$ is replaced by a 
``corresponding" initial state ket $J|\phi_L\ra$ that can be identified with a
state in ${\cal H}_R$. Here $J=CTP^1$ is the operator of 
charge conjugation, time reversal, and reflection across the
Rindler plane, which is a symmetry of all Lorentz invariant quantum
field theories.\footnote{For a configuration eigenstate of a real field, 
the ket $J|\phi_L\ra$ can just be identified with the same function $\phi_L$, 
reflected by an operator $P^1$ 
across the Rindler plane. More generally, $J$ includes 
CT to undo the conjugation of the $\la bra|\rightarrow |ket\ra$ duality.}

The vacuum wave-functional (\ref{Psi0}) can also be represented
as a vector in the Hilbert space ${\cal H}_L\otimes {\cal H}_R$,
by multiplying the amplitudes (\ref{Psi0}) by the corresponding kets
and integrating over the fields:
\bea
|0\ra &\propto& \int {\cal D}\phi_L\, {\cal D}\phi_R\; |\phi_L\ra
|\phi_R\ra\;  \la\phi_R|e^{-\pi H_{\eta}}J|\phi_L\ra\\
&=& \int {\cal D}\phi_L\; |\phi_L\ra
e^{-\pi H_{\eta}}J|\phi_L\ra\\
&=&\sum_n e^{-\pi E_n}|n\ra_L |\bar{n}\ra_R.\label{vac0}
\eea
In the last line the state is expressed in terms of 
eigenstates $|n\rangle$
of the boost Hamiltonian with boost energy $E_n$
(with additional implicit quantum numbers).
It is obtained via $J|\phi_L\ra=\sum_n J|n\ra\la n|\phi_L\ra= \sum_n \la\phi_L|n\ra J |n\ra$, 
using the anti-linearity of $J$. Then the integral over $\phi_L$ yields the identity operator, 
and the result follows since $H_\eta$ commutes with $J$. 
The state $|\bar{n}\ra$ stands for the ``antiparticle state" $J|n\ra$.

This exhibits the precise sense in which 
the quantum field degrees of freedom in the left and right Rindler wedges 
are entangled in the vacuum state.\index{entanglement} 
This entanglement is the origin 
of the correlations between the Hawking quanta and their
partners, and it produces the entanglement entropy\index{entanglement entropy} 
for quantum fields outside a horizon.
Tracing over the state in the
left wedge we obtain the reduced density matrix for the state restricted to
right wedge, 
\beq
\rho_R = {\rm Tr}_L|0\ra\la0|\propto\sum_n e^{-2\pi E_n}|n\ra\la n|.
\eeq
This is the canonical thermal state (\ref{rhoR}) mentioned 
above.\footnote{Its matrix elements could also have been obtained directly 
using the wave functional (\ref{Psi0}), via
$\int {\cal D}\phi_L\; \Psi_0[\phi_L,\phi_R]\Psi_0^*[\phi_L,\phi_R']
\propto\la\phi_R|e^{-2\pi H_{\eta}}|\phi_R'\ra$.}
The horizon entanglement entropy\index{horizon entanglement entropy}
 is the entropy of this thermal state. It diverges as the horizon area times
 the square of the momentum cutoff.
 

\section{Hawking effect}\index{Hawking radiation}
\label{sec:3}

The essence of the Hawking effect~\cite{Jacobson:2003vx} is that the 
correlated vacuum fluctuations described in the
previous section  exist near the horizon of 
a black hole, which is locally equivalent to a 
Rindler horizon. The crucial difference from flat space is that
tidal effects of curved spacetime
peel apart the correlated partners.
The outside quanta sometimes escape to infinity and sometimes
fall backwards into the black hole, while the inside ones
fall deeper into the black hole. 
The escaping quanta have a thermal spectrum 
with respect to the analogue of the boost Hamiltonian,
that is, with respect to the Hamiltonian for the 
horizon-generating symmetry. If the horizon generating Killing
vector is normalized to have unit surface gravity, like
the boost Killing vector, the temperature is again
the Rindler temperature $T_R=\hbar/2\pi$ (\ref{TR}).
However, for a quantum that
escapes from the black hole region, the natural definition
of energy is the generator of asymptotic time translations.
For defining this energy 
we normalize the time translation Killing vector at infinity.
Then the black hole horizon has a surface gravity $\kappa$,
and the temperature is the Hawking temperature,\index{Hawking temperature}
\beq\label{T_H}
T_H=\hbar \kappa/2\pi.
\eeq
Note that the Unruh temperature (\ref{T_U}) 
can be expressed in exactly the
same way as the Hawking temperature since, 
as explained in Sec.\ \ref{sec:kappa}, when the boost Killing field 
is normalized to unity on a given hyperbola the surface gravity
of the Rindler horizon is precisely the acceleration of that hyperbola. 

For a rotating black hole, as explained in Sec.\ \ref{sec:KillingH}, 
the horizon generating Killing vector 
is $\partial_t+\Omega_H\partial_\phi$. The eigenvalues of 
the Hamiltonian 
corresponding to this Killing vector are\footnote{The
sign of the $L$ term is opposite to that of the $E$ term because 
$\partial_\phi$ is spacelike while $\partial_t$ is timelike.} $E-\Omega_HL$,
where $E$ and $L$ are the energy ``at infinity" 
and angular momentum respectively. Thus the Boltzmann factor
for the Hawking radiation is $e^{-(E-\Omega_H L)/T_H}$.
The angular velocity $\Omega_H$ plays the role of a chemical
potential for the angular momentum. 

Missing from this explanation of the Hawking effect 
is the specification of the incoming state.
In principle, there are two places where the state can ``come in" from: spatial
infinity, and the horizon. The state coming from the horizon is determined to be the 
local vacuum by a regularity condition, since anything other than the vacuum
would be singular as a result of infinite blueshift 
when followed backwards in time toward the horizon. 
This is what accounts for the universality of the thermal emission.
However the state coming in from 
infinity has freedom. If it is the vacuum, the
state is called the ``Unruh state",\index{Unruh state} while if it is a thermal state,
as approprate for thermal equilibrium of a black hole with its surroundings, 
it is the ``Hartle-Hawking" state.\index{Hartle-Hawking state}
In the neighborhood of the intersection of past and future horizons,
the Hartle-Hawking state is close to the local Minkowski vacuum. 

For black holes in general relativity, the above description of the
Hawking effect is, in a sense, the complete story. 
For analogue models, however, 
one wants a derivation that does not assume Lorentz invariance,
and that shows the way to the modifications brought about by the
lack thereof.
Also, it is important to be able to allow for experimental conditions 
that determine different incoming states. Moreover, in
the analogue case the horizon state need not be the vacuum, since
in the presence of Lorentz violating dispersion a different state 
can exist without 
entailing anything singular on the horizon. 
Thus we now take a very different viewpoint, 
analyzing the vacuum ``mode by mode". It is this approach that 
Hawking originally followed when he discovered black hole
radiation. It should be emphasized at the outset however that, unlike the 
previous treatment, this approach will apply only to free field theory,
with uncoupled modes satisfying a linear field equation.
 
\subsection{Mode solutions}
My aim here is to convey the essence of the Hawking effect,
using a language that is easily adapted to analogue models in 
which dispersive effects play a role. Hence I will discuss only
a system with one spatial  dimension, and will highlight the
role of the dispersion relation, using WKB methods.

Consider a scalar field $\varphi$ that satisfies
the wave equation $\del_\a(\sqrt{-g}g^{\a\b}\del_\b\vphi)=0$. 
For the metric (\ref{1+1}) we have $\sqrt{-g} = c$ and 
$g^{tt}=1/c^2$, $g^{tx}=v/c^2$, $g^{xx}=(v^2-c^2)/c^2$.
Since the metric is independent of $t$ we can find solutions
with definite Killing frequency, $\vphi=e^{-i\o t}u(x)$. 
Because of the redshift effect an outgoing solution has very rapid 
spatial oscillations of $u(x)$ near the horizon. We can 
thus find an approximate solution
near the horizon by neglecting all terms in which there 
is not at least one derivative of $u(x)$. This yields the equation
\beq\label{u1}
\del_x[(v^2/c-c)\del_x u]=(2i\o v/c)\del_xu.
\eeq
Near a horizon $x=x_H$ where $v=-c$ we have the expansions
$v/c=-1 + O(x-x_H)$ and $v^2/c-c=-2\k (x-x_H) + O[(x-x_H)^2]$. 
Thus at the lowest order in $x-x_H$ the 
near horizon approximation of (\ref{u1}) becomes
\beq\label{u2}
\del_x[(x-x_H)\del_x u]=(i\o/\k)\del_xu, 
\eeq
whose solutions have the form
\beq\label{u3}
u\sim (x-x_H)^{i\o/\k}=e^{i(\o/\k)\ln (x-x_H)}.
\eeq
The logarithmic divergence in the phase justifies the
dominance of spatial derivatives of $\vphi$ near the
horizon. Note that this mode has the same form 
as (\ref{log}), which we inferred in Sec. \ref{redshift} using the
equation of outgoing lightrays to propagate the phase of the wave in the
near horizon region.

Now let's see how to arrive at the same approximate
solution using the 
dispersion relation with the fluid picture. 
First, a mode solution in a 
homogeneous fluid has the 
form $\varphi\sim e^{-i\o t}e^{ikx_{}}$, where 
$x_{}$ is the position in the fluid frame and 
the dispersion relation is $\o^2=F(k)^2$ for some 
function $F(k)$. For instance, for a nondispersive
wave with speed $c$ we have simply $F(k)=ck$. 
If the fluid is flowing with speed $v$ relative to 
the ``lab" then $x=x_f + vt$, where $x_{f}$
is at rest with respect to the fluid. 
In terms of $x_f$ the mode is 
$e^{-i(\o-vk) t}e^{ikx_f}$, which allows us
to read off the frequency as measured in the fluid frame,
$\o_f = \o - vk$. The dispersion relation holds in the 
fluid frame, so we have $\o - vk=\pm F(k)$.

If the flow velocity $v(x)$ is not uniform,
$\o_f = \o - v(x)k$ is \emph{locally} accurate
provided the change of $v(x)$ over a wavelength is 
small compared to $v(x)$ itself. 
The local dispersion relation then becomes
\beq
\o - v(x)k=\pm F(k),
\eeq
which for a fixed 
Killing frequency yields a position-dependent wavevector,
$k_\o(x)$. It should be emphasized that the Killing frequency
$\o$ is a well-defined global constant for a solution, even if the Killing
vector is is not everywhere timelike. 

An approximate, WKB mode solution, taking into account 
only the phase factor, is thus
\beq
u(x)\sim \exp\left({i\int^x k_\o(x') dx'}\right).
\eeq
Finally, if the local
wave velocity $c(x)$ also depends on position in the
fluid (but is time independent in the lab frame),
then the function $F(k,x)$ also depends on position.
If $c(x)$ changes slowly over a wavelength, then 
the mode of the same form is again a good approximation.
For the case of relativistic dispersion $F(k,x)=c(x)k$ we obtain 
$k_\o=\o/(c+v)$  for the outgoing mode. Expanding around the
horizon this yields $k_\o(x)=(\o/\k)(x-x_H)^{-1}$, and so the 
mode takes the same form as (\ref{u3}) derived above.

\subsection{Positive norm modes and the local vacuum}

When the field is quantized, the Hilbert space is
constructed as a Fock space built from single particle
states corresponding to (complex) solutions to the field equation 
with positive conserved ``norm". The norm can be identified
using a conserved inner product, the existence of which follows from 
global phase invariance of the action.
Here I will not attempt to explain the details of this 
construction, which can be found in many expositions,\footnote{For a pedagogical
introduction see, e.g.~\cite{Jacobson:2003vx}, or references therein.}   
but instead will try to provide a simple argument that
captures the essence of the story.
In this section the relativistic case
will be explained, and in the last section I will make some
brief comments about what happens when there is Lorentz violating
dispersion for short wavelengths. The quantum field is taken to 
be a hermitian scalar, which arises from quantization of a real scalar field.

Positive norm modes that are localized can be recognized
as those that have positive frequency in the fluid frame. In the relativistic
case, this amounts to positive frequency in any freely falling frame.
The time derivative in the fluid frame is
$(\del_t)_f= \del_t + v\del_x$. For a mode of the form (\ref{u3}) near the
horizon, this is dominated by the second term, and $v\approx -c$,
hence for such modes 
positive frequency with respect to $t$ in the fluid frame 
is the same as positive frequency with respect to $x$.
(There are two minus signs that cancel: $v = -c<0$ at the horizon, 
but the conventional definition of ``positive frequency" is
$\sim e^{-i\o t}$ with $\o>0$ for temporal frequency, 
and $\sim e^{+ik x}$ with $k>0$ for spatial frequency .)

The mode (\ref{u3}) with logarithmic phase divergence 
at the horizon 
can be analytically continued across the horizon 
to make either a positive or a negative frequency solution.
To see how this works, let's first simplify the notation
a bit and set $x_H=0$, so the horizon lies at $x=0$.
Now a positive $x$-frequency function has the form 
$\int_0^\infty dk f(k) e^{ikx}$, which is analytic in the upper-half
complex $x$-plane since addition of a positive imaginary part to 
$x$ leaves the integral convergent. Similarly, a negative 
$x$-frequency function is analytic in the lower half $x$-plane.
The argument of the logarithm is $x=e^{i\theta}|x|$, so 
$\ln x = i\theta + \ln |x|$. Continuing to $-x$ in the upper or lower 
half plane thus gives $(\ln x)_\pm = \pm i\pi + \ln |x|$ respectively, hence
\beq
e^{i(\o/\k)\ln x}\rightarrow e^{\mp\pi\o/\k}e^{i(\o/\k)\ln |x|}.
\eeq
We can thus write down positive and negative frequency
continuations,
\bea \label{qp}
q_+ &=& u + e^{-\pi\o/\k}\, \ut\\
q_- &=& e^{-\pi\o/\k}\, u + \ut,\label{qm}
\eea
where $u=\theta(x) e^{i(\o/\k)\ln x}$
and $\ut=\theta(-x) e^{i(\o/\k)\ln |x|}$, and  $N$ is a normalization factor.
(The negative frequency continuation $q_-$
has been multiplied by $e^{-\pi\o/\k}$ to 
better reflect the symmetry and thus simplify 
the following discussion.)

We can now express $u$ as a superposition
of positive and negative norm parts, 
\beq\label{upum}
u= u_+ + u_- \propto q_+ - e^{-\pi\o/\k} q_-.
\eeq
From the symmetry of the construction, the
norms of $q_+$ and $q_-$ are equal up to a 
sign, hence the ratio of the squared norms 
(denoted $\la,\ra$) of the negative
and positive norm parts of $u$ is
\beq\label{ratio}
\frac{|\la u_-,u_-\ra|}{\la u_+,u_+\ra}= e^{-2\pi\o/\k} = e^{-E/T_H}.
\eeq
In the last equality I've defined the energy $E=\hbar\o$, 
and $T_H=\hbar\kappa/2\pi$ is the Hawking
temperature. This ``thermal ratio" is the signature of the
Hawking effect, as indicated via the mode $u$ outside the
horizon.  Note that this ratio is a property of the classical solution
to the wave equation, and is determined by the ratio of the
frequency to the surface gravity. Planck's constant 
enters only when we express the result 
in terms of the energy quantum $\hbar\o$.
Note also that if the Killing vector is rescaled, then the
Killing frequency $\o$ and surface gravity $\k$ are 
rescaled in the same way, so that the ratio $\o/\k$ is unchanged.

The presence of the negative frequency part 
$u_-$ in $u$ (\ref{upum}) is unexpected from the 
WKB viewpoint. It corresponds to a negative wavevector,
whereas when we solved the local dispersion relation we found 
$k_\o(x)=(\o/\k)(x-x_H)^{-1}$. Since the support of $u$ 
lies outside the horizon at $x>x_H$, it might seem that this
dispersion relation implies that $k_\o(x)$ is positive, and thus that
the frequency is purely positive. However this is a misconception,
because a function with support on a half line cannot
have purely positive frequency. The concept of a definite
local wavevector must therefore have broken down. Indeed,
if we examine the change of $k$ over a wavelength we find 
$(dk/dx)/k \sim (\k/\o) k$, which is \emph{not} much smaller than $k$
unless $\o\gg\k$. This resolves the puzzle.\footnote{However, it 
raises another one: why did the WKB type mode $\sim \exp(i\int^x k_\o(x') dx')$ 
agree so well with the mode function (\ref{u3})? 
The answer is that (\ref{u2}) is a first order
equation, not a second order one, 
once an overall $\partial_x$ derivative is peeled off.}

\subparagraph{The local outgoing vacuum}
The local outgoing vacuum contains no outgoing excitations.
More precisely, it
is the ground state in the Fock space of 
outgoing positive norm modes.
The outgoing modes we have been discussing are not themselves 
localized, but one can form localized wavepackets from 
superpositions of them with different frequencies. Hence 
we may characterize the local outgoing vacuum by the requirement that  
it be annihilated by the annihilation operators\footnote{What I
am calling the annihilation operator  here 
is related to the field operator $\phi$ by $a(f)=\la f,\phi\ra$,
where $f$ is a positive norm mode.
If $f$ is not normalized this is actually $\la f,f\ra^{1/2}$ 
times a true annihilation operator.} $a(q^+)$
and $a(q_-^*)$ for all positive norm modes.

These operators can be expressed in terms of the annihilation and 
creation operators corresponding to $u$ and $\ut$
using (i) linearity, (ii) equations (\ref{qp}) and (\ref{qm}), and (iii)
the relation $a(f)=-a^\dagger(f^*)$
which should be used if $f$ has negative norm.\footnote{The minus sign comes
from the conjugation of a factor of $i$ in the definition of the norm,
which I will not explain in detail here.}
For example, $a(q_+)=a(u)+ e^{-\pi\o/\k}\, a(\ut)=a(u)- e^{-\pi\o/\k}\, a(\ut^*)$.
The vacuum conditions 
\bea\label{vac}
a(q_+)|0\ra&=&0\\
a(q_-^*)|0\ra&=&0 
\eea
thus amount to 
\bea\label{vacaad}
a(u)|0\ra = e^{-\pi\o/\k}\, a^\dagger(\ut^*)|0\ra\\
a(\ut^*)|0\ra = e^{-\pi\o/\k}\, a^\dagger(u)|0\ra.\\
\eea
If we normalize the mode $u$, 
then the commutation relation $[a(u),a^\dagger(u)]=1$ holds
and implies that, in effect, $a(u)=\partial/\partial a^\dagger(u)$,
and similarly for $\ut$. Thus (\ref{vacaad}) can 
be solved to find the vacuum state for these particular modes
of frequency $\o$, 
\beq\label{vac1}
|0\ra \propto \exp\left(e^{-\pi\o/\k}a^\dagger(u)a^\dagger(\ut^*)\right)|0_L0_R\ra,
\eeq
where $|0_L0_R\ra$ is the state with no $u$ or $\ut^*$ excitations on either side of the 
horizon, $a(u)|0_L0_R\ra = 0=a(\ut^*)|0_L0_R\ra$. In flat space $|0_L0_R\ra$ is called
the (outgoing factor of the) ``Rindler vacuum", while in a black hole spacetime it is the ``Boulware vacuum".

Expanding the exponential in (\ref{vac1}) we obtain another expression for the vacuum
\beq\label{vac2}
|0\ra \propto\sum e^{-n\pi\o/\k}|n_Ln_R\ra,
\eeq
where $n_{L}$ and $n_R$ are the number of particles in the given mode.\footnote{Here I've use the relation
$(a^{\dagger})^n|0\ra = \sqrt{n!}|n\ra$.}
Taking the product over all frequencies, we then arrive at an expression for
the local vacuum of a free field theory near the horizon that has the same form 
as the general thermal result (\ref{vac0}) obtained earlier using the path integral.
The results look different only because here the energies of free field states
with $n$ quanta are given by $n\hbar\o$, and because
here the Killing vector is not normalized to unit surface gravity.

\subsection{Stimulated emission of Hawking radiation}\index{stimulated emission}
\label{stim}

So far I spoke only of the Hawking effect arising from the local vacuum at the horizon.
For a real black hole this is probably the only relevant condition, but for analogue
models it is possible, and even unavoidable because of thermal fluctuations, noise,
or coherent excitations, that the in-state is {\it not} the vacuum. Then
what arises is stimulated emission of Hawking radiation~\cite{Wald:1976ka},
just as the decay of an excited atomic state can be stimulated by the presence of a photon.

To quantify this process, instead of imposing the vacuum  condition (\ref{vac}) 
we can assume the quantum field is in an excited state, 
 \bea
 a^\dagger(\hat q_+)a(\hat q_+)|\Psi\ra&=&n_+|\Psi\ra \\
 a^\dagger(\hat q_-^*)a(\hat q_-^*)|\Psi\ra&=&n_-|\Psi\ra,
\eea
where the $\hat q_\pm$ are normalized versions of (\ref{qp},\ref{qm}).
A simple way to diagnose the emission is via  the expectation value 
of the occupation number of the normalized mode $u$. Using (\ref{upum}) and (\ref{ratio}) we find
\bea
\la\Psi|a^\dagger(u)a(u)|\Psi\ra &=&\la\Psi|a^\dagger(u_+)a(u_+)+ a(u_-^*)a^\dagger(u_-^*)|\Psi\ra \\
&=&\la u_+,u_+\ra[n_+ + e^{-2\pi\o/\k}(n_-+1)]\\
&=&n_+ + \frac{n_+ + n_- +1}{e^{2\pi\o/\k} -1}
\eea
where $\la u_+,u_+\ra=1/(1-e^{-2\pi\o/\k})$. Thus both $n_+$ and $n_-$ stimulate
Hawking emission, while only $n_+$ shows up in the non-thermal spectrum.
Had the state been a coherent state, the occupation  numbers would be replaced
by squared amplitudes. Something analogueous to this occurs in the surface wave 
white hole radiation experiments~\cite{Weinfurtner:2010nu}, although those waves
do not have a relativistic dispersion relation. In the case of a Bose condensate,
the appropriate in-state would presumably be more like a thermal state~\cite{Macher:2009nz}.

\section{The trans-Planckian question}\index{trans-Planckian question}

The sonic black hole was originally conceived by Unruh~\cite{Unruh:1980cg}   in part
to address what has come to be called the trans-Planckian question:
Can the derivation of Hawking radiation be considered reliable given that 
it refers to arbitrarily high frequency field modes? If one assumes local
Lorentz invariance at arbitrarily large boosts, then any high frequency 
mode can be Doppler shifted to low frequency, so one might argue that 
there is nothing to be concerned about. Sometimes the point is raised
that there is an arbitrarily large invariant center of mass energy in the collision between
ingoing and outgoing modes in the vacuum outside a horizon. However, this is true even in flat spacetime.
We never see the effects of such collisions because they concern the ``internal structure"
of the ground state. We could presumably see this quantum gravity 
structure of the vacuum only with probes that have Planckian invariant energy. 
Hence it is not clear to me that there is anything to worry about in the derivation,
provided one is willing to assume local Lorentz symmetry at boost factors arbitrarily 
far beyond anything that will ever be tested. 

Even without assuming exact Lorentz symmetry,
one can infer the Hawking effect by assuming that the outgoing modes are in
their local ground state near the horizon for free-fall frequencies high compared to,
say, the light-crossing time of the black hole, but small compared to the Planck frequency~\cite{Jacobson:1993hn}.
Validity of this assumption is highly plausible since the black hole formation, and 
field propagation in the black hole background, is very slow compared to frequencies
much higher than the light crossing time. One would thus expect that whatever is 
happening in the vacuum, it remains unexcited, and the outgoing modes would emerge
in their ground state in the near horizon region.
The sonic model and other analogues allow this hypothesis to be tested in well-understood 
material systems that break Lorentz symmetry. 

Thus
one is led to consider Hawking radiation in the presence of high frequency/short wavelength dispersion,
both because of the possibility that spacetime is Lorentz violating (LV), and because of the
fact that analogue models are LV. However, given the very strong observational
constraints on Lorentz violation~\cite{Mattingly:2005re},  
  as well as the difficulty of accounting 
for low energy Lorentz symmetry in a theory that is LV in the UV~\cite{Collins:2006bw},   
the possibility of fundamental LV seems rather unlikely. Hence
the main motivations for considering LV dispersion are to understand condensed matter 
analogues, and to have an example---probably unphysical from a fundamental 
viewpoint---in which the vacuum has strong UV modifications and the existence of 
Hawking radiation can be checked. 

The central issue in my view is the origin of the outgoing modes~\cite{Jacobson:1996zs}.
In a condensed matter model with a UV cutoff these must arise from somewhere
other than the near horizon region, either from ``superluminal" modes behind the horizon,
from ``subluminal" modes that are dragged towards the horizon and then released, 
or from no modes at all. The last scenario refers to the possibility that modes
``assemble" from microscopic degrees of freedom in the near horizon region.
This seems most likely the closest to what happens near a spacetime 
black hole, and for that reason deserves to be better understood. Other than
a linear model that has been studied in the cosmological context~\cite{Parentani:2007uq},
and a linear model of quantum field theory on a 1+1 dimensional growing lattice~\cite{Foster:2004yc},
I don't know of any work focusing on how to characterize or study such a process. 

\section{Short wavelength dispersion}\index{dispersion}

In this concluding section, I discuss what becomes of
the Hawking effect when the dispersion relation is Lorentz invariant (``relativistic") for 
long wavelengths but not for short wavelengths, as would be relevant for
many  analogue models. First I summarize results on the robustness of the
``standard" black hole radiation spectrum, and then I describe the phenomena 
of stimulated emission and white hole radiation. 

Dispersion relations of the form 
$\omega^2 = c^2 (k^2 \pm k^4/\Lambda^2)$ have been exhaustively studied.
The plus sign gives ``superluminal" propagation at high wavevectors, while the
minus sign gives ``subluminal" propagation. 
Roughly speaking, a horizon (for long wavelengths) 
will emit thermal Hawking radiation in a given mode provided 
that there is a regime near the horizon in which the mode is relativistic
and in the locally defined vacuum state. This much was argued carefully in Ref.~\cite{Jacobson:1993hn},
and much subsequent work has gone into determining the precise conditions under
which this will happen, and the size of the deviations from the thermal spectrum,
for specific types of dispersion relations. The dispersion determines how the outgoing
modes  arise, that is whether they come from inside or outside the horizon, 
and what quantum state they would be 
found in if the initial state were near the ground state of the field, as in Hawking's 
original calculation. 

The most recent and most complete analysis of the effects of dispersion on the spectrum
can be found in Ref.~\cite{Coutant:2011in},  
in which many references to earlier work can also be found. The basic technique
used there is that of matched asymptotic expansions, pioneered in Refs.~\cite{Brout:1995wp,Corley:1997pr}
as applied to Hawking radiation for dispersive fields. The dispersive modes have 
associated eikonal trajectories with a turning point outside or inside the horizon 
for the sub- and super-luminal cases respectively. Away from 
the turning point approximate solutions can be found using WKB methods. If the 
background fluid velocity (or its analogue) has a linear form $v(x)= -1 +\kappa x $ 
to a good approximation out beyond the turning point, then one can match a
near horizon solution to WKB solutions, and use this to find the Hawking radiation 
state and correlation functions. The near horizon solution is most easily found 
in $k$ space, because while the mode equation is of higher order in $x$ derivatives,
$v(x) = -1 + i\kappa \partial_k$ is linear in $k$ derivatives, so the mode equation is 
second order in $\partial_k$. Further simplifications come about because a linear
$v(x)$ in fact corresponds to de Sitter spacetime, which has an extra symmetry that 
produces factorized modes. One factor is independent of the dispersion and  
has a universal $\omega$ dependence, while the other factor is independent of 
$\omega$ and captures the dispersion dependence. 

The result, for dispersion relations
of the form $\omega^2 = c^2 (k \pm k^{2n+1}/\Lambda^{2n})^2$ (chosen for convenience
to be a perfect square), is that the relative deviations
from the thermal spectrum are no greater than of order 
$(\kappa/\Lambda)(\kappa x_{\rm lin})^{-(1+1/2n)}$
times a polynomial in $\o/\k$.\footnote{For frequencies of order the surface gravity,
this quantity can also be expressed as
$(x_{\rm tp}/x_{\rm lin})^{1+1/2n}$, where $x_{\rm tp}$ is the ($\o$-dependent) WKB turning point.} 
Here the horizon is at $x=0$, and $x_{\rm lin}$ is the largest $x$ for which $v(x)$ has the linear form to a
good approximation.
Thus while it is important that the Lorentz violation wavevector scale $\Lambda$ be 
much greater than the surface gravity $\k$, this may not be good enough to ensure agreement
with the relativistic Hawking spectrum if the linear regime 
of the velocity extends over a distance much shorter than the inverse surface gravity. 

At the other extreme, when the surface gravity is much larger than the
largest frequency for which the turning point falls in the linear region, 
the spectrum of created excitations has been found to be
proportional to $1/\o$, at least for dispersion relations of the form 
$\omega^2 = c^2 (k^2 \pm k^4/\Lambda^2)$.
This is the
low frequency limit of a thermal spectrum, but the temperature is set not
by the surface gravity but by $\sim \Lambda(\kappa x_{\rm lin})^{3/2}$.
This result applies even in the limit of an abrupt ``step" at which the velocity 
changes discontinuously from sub- to supersonic~\cite{Macher:2009tw,Finazzi:2012iu}. 

\subsection{Stimulated Hawking radiation and dispersion}\index{stimulated emission}

For a relativistic free field, the ancestors of Hawking quanta can be traced
backwards in time along the horizon to the formation of the horizon, and then out to infinity.
They are thus exponentially trans-Planckian. In the presence of dispersion,  
blueshifting is limited by the scale of dispersion, so that ancestors can be traced 
back to incoming modes with wave vectors of order $\Lambda$. If the dispersion is 
subluminal, those modes come from outside the black hole horizon, while if
it is superluminal, they come from behind the horizon. Either way, they are
potentially accessible to the control of an experiment. Instead of being in 
their ground state, they might be intentionally 
populated in an experiment, or they might be inadvertently thermally populated. Either
way, they can lead to stimulated emission of Hawking radiation, as discussed in 
Section \ref{stim}.  

This opportunity to probe the dependence of the emitted radiation on the incoming state
is useful to experiments, and it can amplify the Hawking effect, making it easier to detect. 
Note however that when the Hawking radiation is stimulated rather than spontaneous,
it is less quantum mechanical, and if the incoming mode is significantly populated 
it is essentially purely classical. 

\subsection{White hole radiation}\index{white hole}\index{white hole radiation}

A white hole is the time reverse of a black hole. Just as nothing can escape from a
black hole horizon without going faster than light, nothing can {\it enter} a white 
hole horizon without going faster than light. 
Einstein's field equation is time reversal
invariant, so it admits white hole solutions. In fact 
the Schwarzschild solution is 
time reversal symmetric: when taken in its entirety it includes a white hole.
A black hole that forms from collapse is of course not time reversal invariant,
but the time reverse of this spacetime is also a solution to Einstein's equation. 
It is not a solution we expect to
see in Nature, however, both because we don't expect the corresponding initial 
condition to occur, and because, even if it did, the white hole would be 
gravitationally unstable to forming a black hole due to accretion of matter~\cite{Eardley:1974zz, Barrabes}.
Moreover, even if there were no matter to accrete, the horizon would be classically and 
quantum mechanically unstable due to an infinite blueshift effect, as will be explained below.

White hole analogues, on the other hand, can be engineered in a laboratory, 
and are amenable to experimental investigation. For example, one could be realized
by a fluid flow with velocity decreasing from supersonic to subsonic in the direction of the flow. 
Sound waves propagating against the flow would slow down and blueshift as they
approach the sonic point, but the blueshifting would be limited by short wavelength dispersion,
so the white hole horizon might be stable. 
If the horizon is stable, then the time reverse of the Hawking effect will
take place on a white hole background, and the 
emitted radiation will be thermal, at the Hawking temperature of the white hole horizon~\cite{Macher:2009tw}
(see also Appendix D of Ref.~\cite{Macher:2009nz}).
Underlying this relation is the fact that the modes on the white hole background are the time reverse of 
the modes on the time-reversed black hole background. Note that this means that the roles of 
the in and out modes are swapped. In particular, the incoming vacuum relevant to the Hawking radiation
consists of low wavenumber modes propagating against the flow. 

When such a mode with positive 
norm approaches the white hole horizon, it is blocked and begins blueshifting. 
At this stage, it has become a superposition of positive and negative co-moving frequency (and therefore
positive and negative norm) parts. If it were relativistic at all scales, it would continue blueshifting
without limit. It would also be unentangled with the other side of the horizon, so would evidently be 
in an excited state, not the co-moving ground state. Hence there would be a quantum instability
of the vacuum in which the state becomes increasingly singular on the horizon. A classical perturbation
would behave in a similarly unstable fashion.

In the presence of dispersion, however, the blueshifting is arrested when the 
it reaches the dispersion scale. At that stage,
if the mode becomes superluminal, it accelerates and both parts propagate across the horizon. If instead it
becomes subluminal, then it slows down and both parts get dragged back out with the flow. In either case,
the positive and negative norm parts are in an entangled, excited state that is thermal when tracing
over one of the pair. Thus, a dispersive wave field exhibits Hawking radiation from a white hole horizon,
but with two marked differences when compared to black hole radiation: the 
Hawking quanta have high wavevectors even when the Hawking temperature is low, and the entangled
partners propagate on the same side of the horizon (inside for superluminal, outside for subluminal dispersion). 
While on the same side, the partners can separate, since in general they have different group velocities. 

There is an important potential complication with this story of white hole radiation. Although the 
singularity that would arise in the relativistic case is cured by dispersion, an avatar of it emerges in
the form of a zero Killing frequency standing wave. This has been shown to arise from the 
zero frequency limit of the Hawking radiation. In that limit, the emission rate diverges
as $1/\omega$, leading to a state with macroscopic occupation number that grows in 
time~\cite{Mayoral,Coutant:2011in}. 
This process can also be seeded by 
classical perturbations, and it grows until nonlinear effects saturate the growth. 
The resulting standing wave, which is a well-known phenomenon in other contexts,
is referred to in the white hole setting as an ``undulation".\index{undulation}
It is composed of short wavelengths that are well into the dispersive regime. Depending on the 
nature of the flow and the saturation mechanism, it could disrupt the flow and prevent a smooth
horizon from forming.

To conclude, I will now describe what was seen in the Vancouver\index{Vancouver experiment}  
experiment~\cite{Weinfurtner:2010nu}.\index{Vancouver experiment} 
That experiment involved a flow of water in a flume tank with a velocity profile that produced a white
hole horizon for long wavelength, shallow water, surface waves (which are dispersionless over a uniform bottom). 
When blueshifted those waves convert to deep water waves, with a lower group velocity, which behave like
the ``subluminal" case described above. In the experiment coherent, long waves with nine different frequencies
were launched from downstream, propagating back upstream towards the
white hole horizon, and the resulting conversion to short waves was observed. 
The squared norm ratio of the negative and positive norm components of the 
corresponding frequency eigenmode  
was consistent with the thermal ratio (\ref{ratio}).\footnote{The relevant norm is 
determined by the action for the modes. This has been worked out assuming irrotational
flow~\cite{Unruh:2012ve}, which is a good approximation although the flow does 
develop some vorticity. The predicted Hawking temperature was estimated, but it 
is difficult to evaluate accurately because of the presence of the undulation,
the fact that it depends on the flow velocity field that was not precisely measured, and
the presence of vorticity which has not yet been included in an effective metric description.}
This can be understood as coherently stimulated emission of Hawking radiation
(see Appendix C of Ref.~\cite{Macher:2009nz} for a general discussion of this process).
It is strictly classical, but it is governed by 
the same mode conversion amplitudes that would produce spontaneous emission if the
system could be prepared in the ground state.  

\section{Acknowledgments}
I am grateful to Renaud Parentani for many helpful discussions on the material presented here,
as well as suggestions for improving the presentation. Thanks also to Anton de la Fuente 
for helpful discussions on the path integral derivation of vacuum thermality. 
This work was supported in part by the National Science Foundation under 
Grant Nos. NSF PHY09-03572 and NSF PHY11-25915.

\bibliographystyle{plain}

\end{document}